\pdfoutput=1
\documentclass[
 reprint,
superscriptaddress,
groupedaddress,
nofootinbib,
 amsmath,amssymb,
 aps,
 pra,
]{revtex4-1}
\usepackage{graphicx}% Include figure files
\usepackage{bm}% bold math
\usepackage{hyperref}% add hypertext capabilities
\usepackage[mathlines]{lineno}% Enable numbering of text and display math
\usepackage[utf8]{inputenc}
\usepackage{amsmath,amssymb}
\usepackage[T1]{fontenc} % Use 8-bit encoding that has 256 glyphs
\usepackage{microtype} % Slightly tweak font spacing for aesthetics
\usepackage[english]{babel} % Language hyphenation and typographical rules

\usepackage{booktabs} % Horizontal rules in tables

\usepackage{graphicx}
\usepackage{amssymb}
\usepackage{braket,mleftright}
\usepackage{empheq}
\usepackage{subfigure}
\usepackage{stackrel}
\usepackage{soul}
\usepackage{blkarray}
\usepackage{multirow}
\usepackage{amsmath}
\usepackage{physics}
\usepackage{amsfonts}
\usepackage{bm}
\usepackage{bbold} %math& text
\usepackage{color}
\newcommand{\beq}{\begin{equation}}
\newcommand{\eeq}{\end{equation}}
\newcommand{\bse}{\begin{subequations}}
\newcommand{\ese}{\end{subequations}}
\newcommand{\bea}{\begin{eqnarray}}
\newcommand{\eea}{\end{eqnarray}}
\usepackage[utf8]{inputenc} %acentos
\usepackage{amsmath}

\usepackage{float}

\usepackage{enumitem} % Customized lists
\setlist[itemize]{noitemsep} % Make itemize lists more compact
\usepackage{hyperref} % For hyperlinks in the PDF
\usepackage{braket}
\usepackage{verbatim} %multiline comments

\begin{document}

%\preprint{APS/123-QED}

\title{Sensing quantum chaos through the non-unitary geometric phase}

\author{Nicol\'{a}s Mirkin}
\email[Corresponding author:]{\,mirkin@df.uba.ar}
\affiliation{%
Departamento de F\'{i}sica “J. J. Giambiagi” and IFIBA, FCEyN, Universidad de Buenos Aires, 1428 Buenos Aires, Argentina
}%

\author{Diego Wisniacki}

\affiliation{%
Departamento de F\'{i}sica “J. J. Giambiagi” and IFIBA, FCEyN, Universidad de Buenos Aires, 1428 Buenos Aires, Argentina
}%

\author{Paula I. Villar}

\affiliation{%
Departamento de F\'{i}sica “J. J. Giambiagi” and IFIBA, FCEyN, Universidad de Buenos Aires, 1428 Buenos Aires, Argentina
}%

\author{Fernando C. Lombardo}

\affiliation{%
Departamento de F\'{i}sica “J. J. Giambiagi” and IFIBA, FCEyN, Universidad de Buenos Aires, 1428 Buenos Aires, Argentina
}%

\date{\today}% 

\begin{abstract}
Quantum chaos is usually characterized through its statistical implications on the energy spectrum of a given system. In this work we propose a decoherent mechanism for sensing quantum chaos. The chaotic nature of a many-body quantum system is sensed by studying the implications that the system produces in the long-time dynamics of a probe coupled to it under a dephasing interaction. By introducing the notion of an effective averaged decoherence factor, we show that the correction to the geometric phase acquired by the probe with respect to its unitary evolution can be exploited as a robust tool for sensing the integrable to chaos transition of the many-body quantum system to which it is coupled. This sensing mechanism is verified for several systems with different types of symmetries, disorder and even in the presence of long-range interactions, evidencing its universality.  
\end{abstract}

\maketitle

%\section{\label{Section-Intro}Introduction}
\section{Introduction} 

Which is the most universal feature of quantum chaos? Is it a spectral or a dynamical property? Historically, quantum chaos was introduced in the literature through a spectral definition, where the chaoticity of a many-body quantum system was characterized through its statistical similarity  to the predictions raised by Random Matrix Theory (RMT) \cite{haake1991quantum,stockmann2000quantum}. %From this point of view, a Wigner-Dyson distribution for the separation of its energy levels was conjectured to be a universal characteristic of a quantum chaotic system and a Poisson distribution to an integrable one. 
However, later on the approach slightly turned to a dynamical definition of quantum chaos. For instance, the seminal work of Peres  proposed the Loschmidt Echo, a measure of irreversibility and sensitivity to perturbations, as a dynamical signature of quantum chaos \cite{peres1984stability}. While Peres original proposal was explicitly conjectured for the long-time regime, since the appearance of some outstanding works that found a link between the short-time behaviour and the Lyapunov exponent \cite{jalabert2001environment,goussev2012loschmidt}, the long-time regime was almost entirely neglected. More recently, the Out-of-Time-Ordered-Correlators (OTOCS) were also proposed as a  dynamical quantifier of quantum chaos \cite{larkin1969quasiclassical,shenker2014black,aleiner2016microscopic,huang2017out} and it has been argued that the long-time regime is the most reliable one for characterizing the chaotic nature of a given many-body quantum system \cite{garcia2018chaos,fortes2019gauging}. Considering that the OTOCS and the Loschmidt Echo are deeply connected \cite{yan2020information}, the question of wether the chaotic nature of a quantum system can be universally measured through the long-time dynamics of the Loschmidt Echo remains unexplored. 

Other remarkable feature of the Loschmidt Echo is its connection with decoherence under pure dephasing scenarios. As a consequence of this, some works have studied the interplay
between chaos and decoherence, for example, by considering quantum systems with a classical chaotic counterpart as a reduced system, usually coupled to regular environments and with the focus on the short-time dynamics \cite{zurek1994decoherence,pattanayak1997exponentially, habib1998decoherence,monteoliva2000decoherence,jalabert2001environment,toscano2005decoherence, wisniacki2009scaling}. On the contrary, the idea of sensing the chaotic nature of an arbitrary many-body environment without classical counterpart through the long-time decoherent dynamics of a probe coupled to it, has not been investigated yet. This is what we call a decoherent characterization of quantum chaos and constitutes the main motivation of our present work. 

We remark that in the past the task of sensing with a probe a certain property of a many-body quantum system has already been demonstrated in the literature, both theoretically and experimentally, for identifying critical points, phase transitions, unknown temperatures and non-Markovian behaviour \cite{quan2006decay,cucchietti2007universal,damski2011critical,haikka2012non}. In those works, a common approach that has already proven its worth is to monitor how the geometric phase acquired by the probe is corrected due to its coupling to the many-body environment with respect to its unitary evolution \cite{fuentes2002vacuum,carollo2005geometric,yi2007geometric,cucchietti2010geometric,martin2013berry,lombardo2020detectable,lombardo2021detectable}. 

\begin{figure}[!htb]
\begin{center}
\includegraphics[width=87mm]{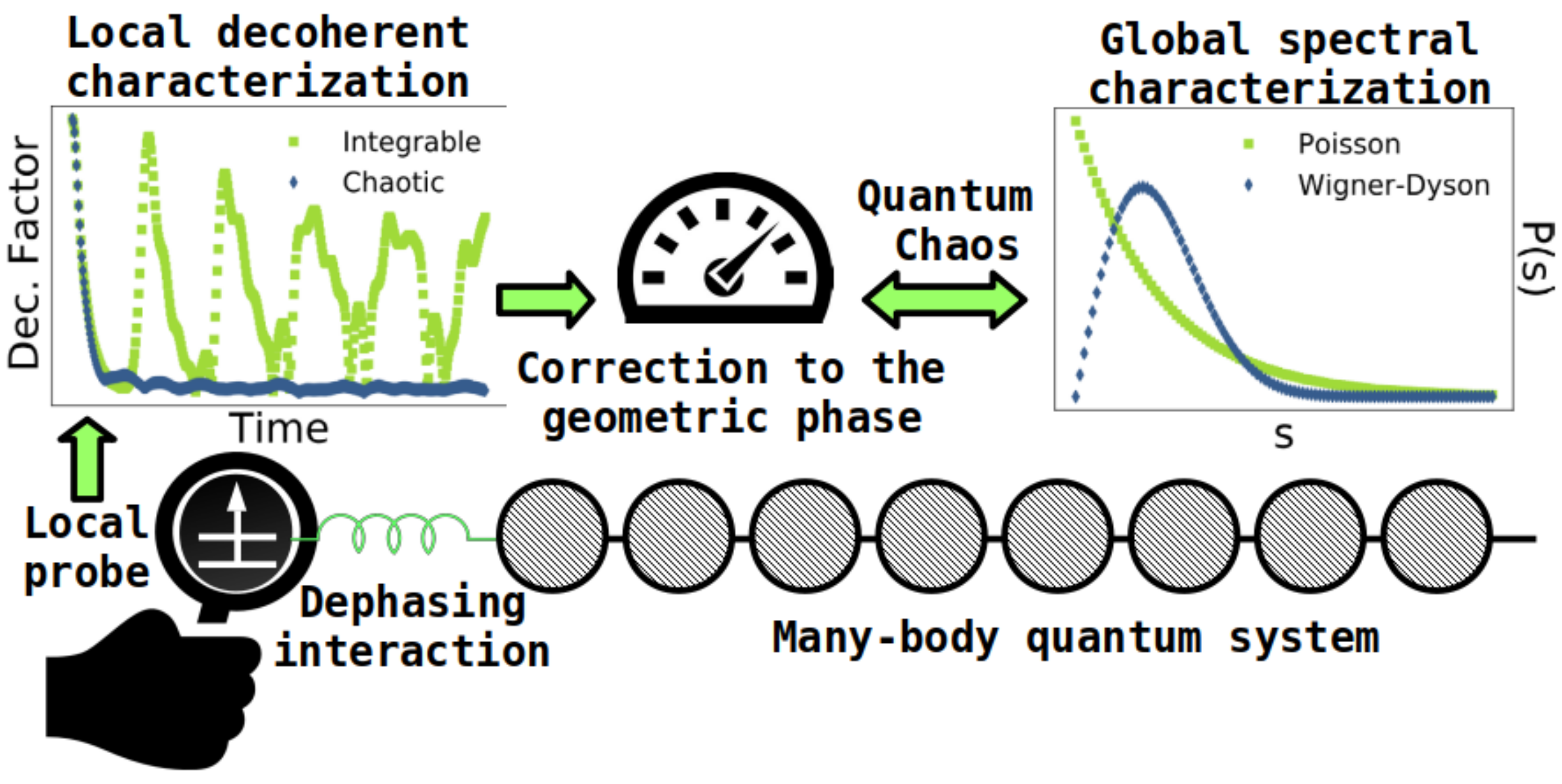}
\begin{footnotesize}
\caption{Schematic representation of our local decoherent characterization of quantum chaos.
The chaoticity of a many-body quantum system is sensed through the long-time decoherent dynamics of a probe coupled to it under a dephasing interaction. While the decoherence factor has huge fluctuations when the energy-level separation of the many-body quantum environment follows a Poisson distribution, almost no coherence revivals are seen in the probe if the statistics is compatible to a Wigner-Dyson distribution. The impact of the decoherence on the probe is quantified through the correction to its accumulated geometric phase with respect to the unitary evolution.} 
\label{fig1}
\end{footnotesize}
\end{center}
\end{figure}

In this work, our ultimate goal is to construct a local decoherent mechanism for sensing quantum chaos. For that purpose, we propose a unified picture that bridges the gap between quantum chaos, decoherence, Loschmidt Echo, long-time dynamics and the non-unitary geometric phase. Our proposal consists on probing the chaoticity of a many-body quantum system by monitoring the decoherence dynamics suffered by a two-level system coupled to the latter through a dephasing interaction (see Fig. \ref{fig1}). Given its robustness and sensitivity, the physical quantity that we use to quantify the detrimental effects generated by the many-body environment to the probe is the correction to its accumulated geometric phase with respect to an isolated unitary evolution. Under this general scheme, we find a precise correspondence between the degree of correction to the geometric phase acquired by the probe and the degree of chaos present in the many-body quantum environment being sensed. Seeking universality, we verify this sensing mechanism for several environmental spin chains with different conserved symmetries, in the presence of disorder and even in the realistic situation of long-range interactions. Remarkably, we achieve the latter by maintaining both the intrinsic Hamiltonian of the probe and the interaction term fixed, which makes our sensing protocol experimentally feasible. Moreover, unlike previous methods considered in the literature, we can reproduce the whole integrable to chaos transition without any consideration about the energy level symmetries and resorting to not so large many-body quantum systems. Despite a similar approach was followed in a recent work \cite{mirkin2021quantum}, there the probe was a reduced part of the chain being sensed and thus the interaction Hamiltonian was not fixed but depended on the specific model under consideration. On the contrary, in this work the physical interpretation underlying the sensing mechanism is more transparent and the quantity that we use to monitor decoherence is more robust for being measured experimentally.    

This work is organized as follows. On Section II we present the general framework under which we built our decoherent characterization of quantum chaos. Subsequently, on Section III we review the standard approach regarding the spectral definition of a quantum chaotic system. We continue on Section IV with a short description about the non-unitary geometric phase. Our main analysis is presented on Section V, where we consider four different physical systems as environments to be sensed. We conclude on Section VI with some final remarks.

\renewcommand{\figurename}{Figure} 

%The role of quantum chaos has been proven as crucial in physical phenomenons such as thermalization of isolated many-body quantum systems, equilibration, information scrambling and so on.   

\section{General framework}
The main idea of our work is schematically summarized on Fig. \ref{fig1}. As we have already posed in the Introduction, our proposal consists on probing the chaotic behaviour of an arbitrary many-body quantum system by monitoring the long-time decoherence dynamics of a two-level quantum probe coupled to it. With this purpose, we will consider a situation where the probe is coupled to the first spin of an environmental chain through a dephasing interaction. The Hamiltonian representing the above setup is given by 

\begin{equation}
\begin{aligned}
\hat{H} & = \hat{H}_S + \hat{H}_{int} + \hat{H}_E \\ &= \frac{\omega}{2} \hat{\sigma}_{0}^{z} + g \hat{\sigma}_{0}^{z}\hat{\sigma}_{1}^{z} + \hat{H}_E   
\end{aligned}
\end{equation}
where $\omega$ is the characteristic frequency of the probe, $g$ the coupling strength to the first spin of the chain and $\hat{H}_E$ the specific environmental Hamiltonian to be sensed. As we are interested in a dynamical approach, we will consider an initial product state of the form
\begin{equation}
    \hat{\rho}(0)= \ket{\psi_{0}}\bra{\psi_{0}} \otimes \ket{\varepsilon(0)} \bra{\varepsilon(0)},
\end{equation}
where the initial state of the probe is
\begin{equation}
\left|\psi_{0}\right\rangle=\cos (\theta / 2)|0\rangle+ e^{i\phi}\sin (\theta / 2)|1\rangle,    
\end{equation}
with $\theta=[0,\pi)$ and $\phi=[0,2\pi)$. Similarly, the initial state of the environment is a pure random state for each spin 
\begin{equation}
\ket{\varepsilon(0)}= \ket{\psi_{1}} \ket{\psi_{2}} \dots \ket{\psi_{L}}     
\end{equation}
where \begin{equation}
\left|\psi_{k}\right\rangle=\cos (\vartheta_k / 2)|0\rangle+e^{i\varphi_k}\sin (\vartheta_k / 2)|1\rangle  \quad \forall k \in [1,L],  
\label{in_state}
\end{equation}
with $\vartheta_k=[0,\pi)$, $\varphi_k=[0,2 \pi)$ and $L$ the number of environmental spins. Let us remark that in order to sense the entire spectrum of the environment in an unbiased way, we must consider several realizations with different random initial pure states. This procedure is important since with a sufficiently large number of realizations, all the eigenstates of the environment will then contribute equally to the dynamics of the probe. Given the dephasing interaction between the probe and the environmental chain, the reduced dynamics of the probe can be solved analytically. Its reduced density matrix is given by 

\begin{equation}
\begin{aligned}
\hat{\rho}_{\mathrm{r}}(t) & =\cos ^{2}(\theta / 2) \ket{0}\bra{0}
+\sin ^{2}(\theta / 2) \ket{1}\bra{1} \\
& +\frac{\sin \theta}{2} e^{- i (\omega t+\phi)}  r(t) \ket{0}\bra{1}
+\frac{\sin \theta}{2} e^{i (\omega t+\phi)}  r^{*}(t)\ket{1}\bra{0},
\end{aligned}
\label{rho_t}
\end{equation}
where $r(t)$ is the so-called decoherence factor and is given by
$r(t)=\left\langle\varepsilon_{1}(t) \mid \varepsilon_{0}(t)\right\rangle$ with $\ket{\varepsilon_{k}(t)}=e^{-i t\left[\hat{H}_{E}+(-1)^{k} \hat{H}_{\mathcal{S}\mathcal{E}} \right]}|\varepsilon(0)\rangle$, where $\hat{H}_{\mathcal{S}\mathcal{E}}$ refers to the term of $\hat{H}_{int}$ acting solely over the environmental degrees of freedom (in our case $\hat{\sigma}_{1}^{z}$). It is important to note that the square of the absolute value of the decoherence factor is exactly what is known as the Loschmidt Echo, which measures the sensitivity of the environment to the specific perturbation induced by its coupling to the probe. 

As has already been posed, we are interested in sensing the environment homogeneously, so we must consider a full mixture of realizations for pure random initial states for the environment ($\hat{\rho}_{\varepsilon_m}(0)=\ket{\varepsilon_m(0)}\bra{\varepsilon_m(0)} \, \text{with} \, m \in [1,R]$). Consequently, if the number of realizations $R$ is sufficiently large, we will be dealing with an effective averaged decoherence factor $\tilde{r}_{e}(t)$ of the form 
%(see Appendix \ref{appA} for more details and a numerical demonstration)
%equal to having considered as an initial state a maximally mixed one $\hat{\rho}_\epsilon (0)=\mathbb{1}/2^L$ (see Appendix \ref{appA} for more details), i.e.
\begin{widetext}
\begin{equation}
\begin{aligned}
    \tilde{r}_{e}(t)&= \lim_{R\to\infty} \left[ \frac{1}{R}\sum_{m=1}^{R}  \bra{\varepsilon_m(0)} e^{i t\left[\hat{H}_{E}- \hat{H}_{\mathcal{S}\mathcal{E}} \right]} e^{-i t\left[\hat{H}_{E}+ \hat{H}_{\mathcal{S}\mathcal{E}} \right]}|\varepsilon_m(0) \rangle \right] \\
    & = \lim_{R\to\infty} \left[ \frac{1}{R}\sum_{m=1}^{R} \Tr \left( \ket{\varepsilon_m(0)}\bra{\varepsilon_m(0)}  %\hat{\rho}_{\varepsilon_m}(0) 
    e^{i t\left[\hat{H}_{E}- \hat{H}_{\mathcal{S}\mathcal{E}} \right]} e^{-i t\left[\hat{H}_{E}+ \hat{H}_{\mathcal{S}\mathcal{E}} \right]} \right) \right]  \simeq \frac{1}{2^L} \Tr \left(  e^{i t\left[\hat{H}_{E}- \hat{H}_{\mathcal{S}\mathcal{E}} \right]} e^{-i t\left[\hat{H}_{E}+ \hat{H}_{\mathcal{S}\mathcal{E}} \right]} \right).  
\end{aligned}
\label{dec_factor_eff}
\end{equation}
\end{widetext}
Due to its averaged nature, this effective decoherence factor is independent of the initial state, containing exclusively the environmental Hamiltonian and its perturbation due to the coupling with the probe. Equivalently, let us note that Eq. (\ref{dec_factor_eff}) is exactly the decoherence factor associated to a maximally mixed initial state for the environment $\hat{\rho}_\epsilon (0)=\mathbb{1}/2^L$ (see Appendix \ref{appA} for more details) and it is also related to the average of the Loschmidt Echo over initial states according to the Haar measure \cite{zanardi2004purity,dankert2005efficient,wisniacki2013sensitivity}. We can clear up Eq. (\ref{dec_factor_eff}) by diagonalizing both evolution operators separately,   

\begin{equation}
\begin{aligned}
& \hat{U}^\dagger = e^{i t\left[\hat{H}_{E}- \hat{H}_{\mathcal{S}\mathcal{E}} \right]} = \sum_{k} e^{it \xi_k} \ket{\xi_k}\bra{\xi_k} \\
& \hat{V} = e^{-i t\left[\hat{H}_{E}+ \hat{H}_{\mathcal{S}\mathcal{E}} \right]} = \sum_{l} e^{-it \eta_l} \ket{\eta_l}\bra{\eta_l},
\end{aligned}
\label{aut_pert}
\end{equation}
where $\set{\xi_k , \eta_l}$ and $\set{\ket{\xi_k},\ket{\eta_l}}$ are the eigenvalues and eigenstates of the operators $\hat{U}^\dagger$ and $\hat{V}$, respectively. Using the above definitions, we have 

\begin{equation}
    \hat{U}^\dagger \hat{V}= \sum_{k=1} \sum_{l=1} e^{-it(\eta_l - \xi_k)} \braket{\xi_k}{\eta_l} \ket{\xi_k}\bra{\eta_l}.
    \label{uv}
\end{equation}
and if we trace over Eq. (\ref{uv}), the effective averaged decoherence factor $\tilde{r}_e (t)$ reduces to

\begin{equation}
\begin{aligned}
    \tilde{r}_e(t)&= \frac{1}{2^L} \Tr \left( \hat{U}^\dagger \hat{V} \right) \\ &
    = \frac{1}{2^L} \left(  \sum_{k=1} \sum_{l=1} e^{-it(\eta_l - \xi_k)} \abs{\braket{\xi_k}{\eta_l}}^2 \right). 
\end{aligned}
\label{r_Tinf}
\end{equation}

This constitutes the universal expression for an averaged decoherence factor associated to a completely arbitrary environment coupled to a probe through a dephasing interaction. In what follows, we will show that the chaotic nature of the environment is hidden in the long-time dynamics of this quantity, involving both the eigenvalues and eigenstates of the perturbed Hamiltonian $\hat{H}_E$. 

\section{Spectral characterization of quantum chaos}
Quantum chaos was first introduced in the literature as a spectral property. Under this context, a many-body quantum system was considered as chaotic if the statistics behind its energy level distribution fulfilled the predictions raised by RMT. Given the statistical nature of this approach, a usual requirement is to work under the limit of high dimensional Hilbert spaces and to separate the energy levels according to their symmetries \cite{percival1973regular, berry1977level,bohigas1984characterization}. By arranging the energy levels in an ordered set $e_n$, we can define the nearest neighbour spacings as $s_n=e_{n+1}-e_n$. Therefore, if the statistics behind  $s_n$ follows a Wigner-Dyson distribution, the quantum system is said to be chaotic. On the contrary, if the statistics is Poissonian, the system is integrable. For quantifying the statistical distance between the actual distribution and a perfectly chaotic or integrable one, it is a standard procedure to use the so-called distribution of $\min(r_n,1/r_n)$, where $r_n$ is the ratio between the two nearest neighbour spacings of a given level ($r_n=s_n/s_{n-1}$). Consequently, the chaotic nature of a given system can be measured through the spectral indicator \cite{oganesyan2007localization,atas2013distribution,kudo2018finite} 

\begin{equation}
\Tilde{r}_n=\dfrac{\min(s_n,s_{n-1})}{\max(s_n,s_{n-1})}=\min(r_n,1/r_n).
\end{equation}
Since the mean value of $\Tilde{r}_n$ ($\overline{\min(r_n,1/r_n)}$) attains a minimum value when the statistics is Poissonian ($\mathcal{I}_{P} \eqsim 0.386$) and a maximum
value when it is Wigner-Dyson ($\mathcal{I}_{WD} \eqsim 0.5307$), we can normalize it as
\begin{equation}
    \eta=\dfrac{\overline{\min(r_n,1/r_n)}-\mathcal{I}_P}{\mathcal{I}_{WD}-\mathcal{I}_P}.
\label{eta}
\end{equation}

Consequently, an integrable regime is characterized by $\eta \simeq 0$ and a chaotic one by $\eta \simeq 1$. We emphasize that this measure is only useful under the limit of large many-body quantum systems such as to have a sufficiently robust energy spectrum to compute the statistics. Also, note that this spectral indicator $\eta$ only depends on the eigenvalues of the spectrum, differently from the averaged decoherence factor on Eq. (\ref{r_Tinf}) which takes into account not only the eigenvalues but the eigenvectors of the system as well.

\section{Non-unitary geometric phase}
During an adiabatic evolution an isolated quantum system can acquire a phase that is geometric in nature, besides its usual dynamical phase \cite{berry1984quantal}. Naturally, if the evolution is non-unitary, the detrimental effects from the environment modify the geometric phase that the open system acquires with respect to its closed evolution. As a consequence of this, great attention has been devoted to understand the robustness of the geometric phase to different types of environment due to its potential implementations on geometric quantum computation \cite{lombardo2006geometric,lombardo2010environmentally,lombardo2013nonunitary,lombardo2015correction,villar2020geometric}. 

The geometric phase acquired by a two-level-system along a unitary cyclic evolution $\hat{H}_S$
is given by 
\begin{equation}
    \Phi_u= N \pi(1+\cos(\theta)),
\end{equation} 
where $N$ is the number of periods under consideration. On the contrary, when the evolution is non-unitary, we can generalize via a quantum kinematic approach the notion of geometric phase along a
quasi-cyclic path $\mathcal{P}: t \in [0,\tau]$ ($\tau= 2\pi N/\omega$) defining 
\begin{equation}
\begin{aligned}
\Phi=\arg \Bigg(\sum_{k=\pm} & \sqrt{\lambda_{k}(0) \lambda_{k}(\tau)}  \left\langle\Psi_{k}(0) \mid \Psi_{k}(\tau)\right\rangle \\ & e^{-\int_{0}^{\tau} d t\left(\Psi_{k}| \frac{\partial}{\partial t}  | \Psi_{k}\right\rangle}\Bigg),
\end{aligned}
\label{eq_phase}
\end{equation}
where $\lambda_k$ are the eigenvalues of the reduced density matrix of the open system and $\ket{\Psi_k}$ its eigenstates \cite{tong2004kinematic}
. This expression is gauge invariant and reduces to the unitary case when the interaction strength between the open system and the environment vanishes ($g=0$). Since under our scheme the initial state of the probe is pure ($\lambda_+(0)=1$ and $ \lambda_-(0)=0$), then Eq. (\ref{eq_phase}) simplifies to 
\begin{equation}
\Phi = \arg \left(\langle\Psi_+(0) \mid \Psi_+(\tau)\rangle \right)-\operatorname{Im} \int_{0}^{\tau} d t\langle\Psi_+ \mid \dot{\Psi}_+\rangle.
\label{fase}
\end{equation}
Thus, we can define the correction of the non-unitary geometric phase $\Phi$ with respect to the unitary case $\Phi_u$ as $\delta \Phi= 1 -\Phi/\Phi_u$.  It is important to remark that the correction to the unitary geometric phase depends strongly on the kind of environment coupled to the main system and that this correction can be measured by means of NMR techniques \cite{cucchietti2010geometric} or by interferometric experiments \cite{zeilinger1988single,leek2007observation,maclaurin2012measurable,wood2020observation}.

\section{Sensing quantum chaos on realistic spin chains}
In this Section we will show our results for the general framework that was previously introduced on Section II, in which a two-level quantum probe is sensing the chaoticity of a realistic spin chain coupled to it under a dephasing interaction. To unveil the degree of quantum chaos present in the environmental chain being sensed, we will monitor how the reduced dynamics of the probe is affected when a certain parameter of $\hat{H}_E$ (that sets the chaoticity of the chain) is swept along some parameter range. As we have already argued, we will study how the geometric phase acquired by the probe along several periods is corrected with respect to its unitary evolution. The reason why we focus in this particular quantity is based on the fact that the non-unitary geometric phase has already proven its worth as a sensor for monitoring other properties of many-body quantum systems. Nevertheless, the task of sensing quantum chaos has been entirely neglected. For the sake of universality, we will test our sensing protocol under the light of four spin chains as different as possible, involving different types of symmetries, disorder and even in the presence of long-range interactions.

\subsection{Ising with transverse magnetic field}
The first environmental spin chain in which we are going to deploy our analysis is an Ising spin chain with transverse magnetic field, whose Hamiltonian is given by
\begin{equation}
 \hat{H}_E = \sum\limits_{k=1}^{L}\left(h_x \hat{\sigma}_{k}^{x}+h_z\hat{\sigma}_{k}^{z} \right) -   J \sum\limits_{k=1}^{L-1}\hat{\sigma}_{k}^{z}\hat{\sigma}_{k+1}^{z},
 \label{ising}
\end{equation}
where $h_x$ and $h_z$ are the strength of the  transverse and longitudinal magnetic field, respectively, $J$ the nearest-neighbor coupling and $L$ the number of spins in the chain. In this model, the parity is a conserved quantity and we must take it into account for studying the integrable to quantum chaos transition with the standard spectral measures. The parity is defined through the  permutation operators $\hat{\Pi}=\hat{P}_{1,L}\hat{P}_{2,L-1}\dots \hat{P}_{L/2-1,L/2+1}$ for a chain of odd length $L$ and for the even case its analogous. This implies that the spanned space is divided into odd and even subspaces with dimension $D=D^{even}+D^{odd}$  ($D^{even/odd}\approx D/2$). While this model is integrable in the limit of $h_z \gg h_x$ and $h_x \gg h_z$, it exhibits quantum chaos when the longitudinal and the transverse field are of comparable strength. As a consequence, the environmental parameter that we will sweep to see how the geometric phase of the probe is affected by the chaoticity of the environmental chain is $h_z$. 

\begin{figure}[!htb]
\begin{center}
\includegraphics[width=87mm]{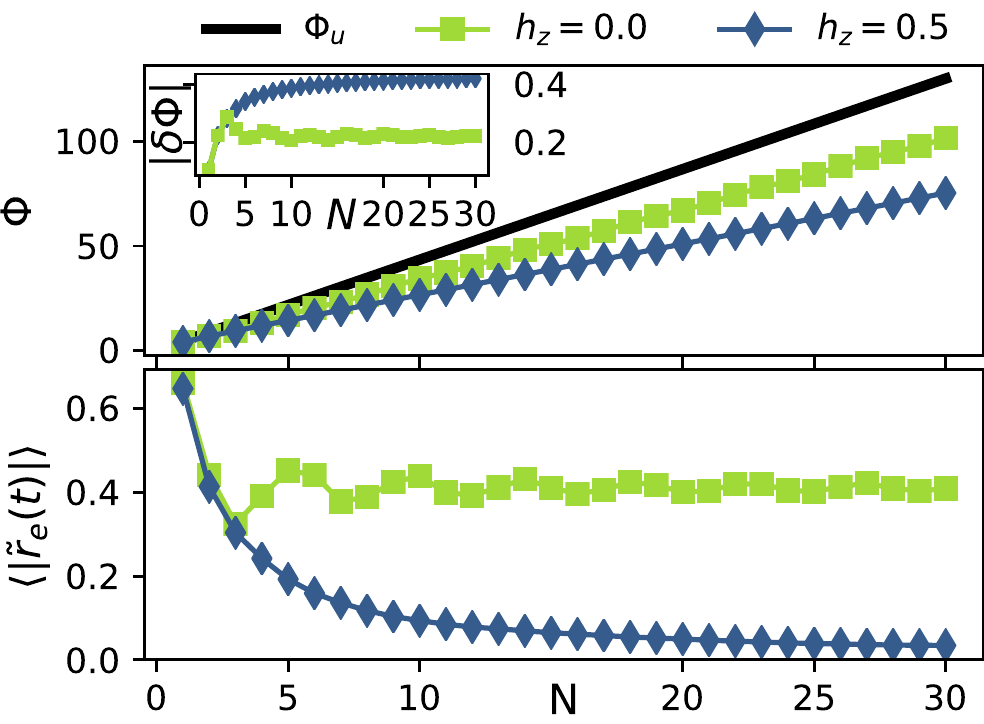}
\begin{footnotesize}
\caption{\textbf{Upper panel:} Total accumulated geometric phase as a function of the number of periods $N$ ($N=\tau \omega/2\pi$)
for a unitary evolution (black solid line), a non-unitary evolution caused by an integrable environmental chain ($h_z=0.0$, green squares) and a chaotic one ($h_z=0.5$, blue diamonds). \textbf{Inset:} Correction to the geometric phase accumulated with respect to the unitary case ($|\delta \Phi|= |1 - \Phi/\Phi_{u}|$) as a function of $N$. \textbf{Lower panel:} Temporal average of the absolute value of $\tilde{r}_e (t)$ as a function of the number of periods $N$ for both integrable and chaotic regimes. In all panels, the initial state of the probe is a fixed pure state $(\theta=3\pi/8; \phi=0)$ and $R=100$ realizations for different pure random initial states for the environment were considered. The rest of the parameters are set as $\omega=1$, $h_x=1$, $J=1$, $g=0.2$, $L=9$.}
\label{fig_dynamics}
\end{footnotesize}
\end{center}
\end{figure}

For an introductory but representative example, in the upper panel of Fig. \ref{fig_dynamics} we show the non-unitary accumulated geometric phase of the probe as a function of the number of periods, for a typical regime where the environment is integrable ($h_z=0 \,;h_x=1$) and for a typical regime where it is chaotic ($h_z=0.5\,;h_x=1$). For greater clarity, in the inset we normalize it with respect to the unitary evolution and there it becomes evident that during the first periods (short-time dynamics), the correction $|\delta \Phi|$ generated by both environmental regimes is indistinguishable. However, as the time evolution grows, the integrable regime leads to a significant lower correction with respect to the chaotic case. This qualitatively different behaviour for $|\delta \Phi|$ can be better understood by comparing with the lower panel of Fig. \ref{fig_dynamics}, where we show the mean value of the averaged decoherence factor $|\tilde{r}_e(t)|$ as a function of the number of periods for the same set of parameters as the inset. While in the chaotic regime the mean value of $|\tilde{r}_e(t)|$ goes to zero at long times, fully destroying the coherences of the probe and thus any vestige of the geometric phase,  in the integrable regime $|\tilde{r}_e(t)|$ oscillates periodically around a mean value much greater than zero (see Fig. 1) and the information about the geometric phase is not totally lost. 

\renewcommand{\figurename}{Figure} 

\begin{figure}[!htb]
\begin{center}
\includegraphics[width=87mm]{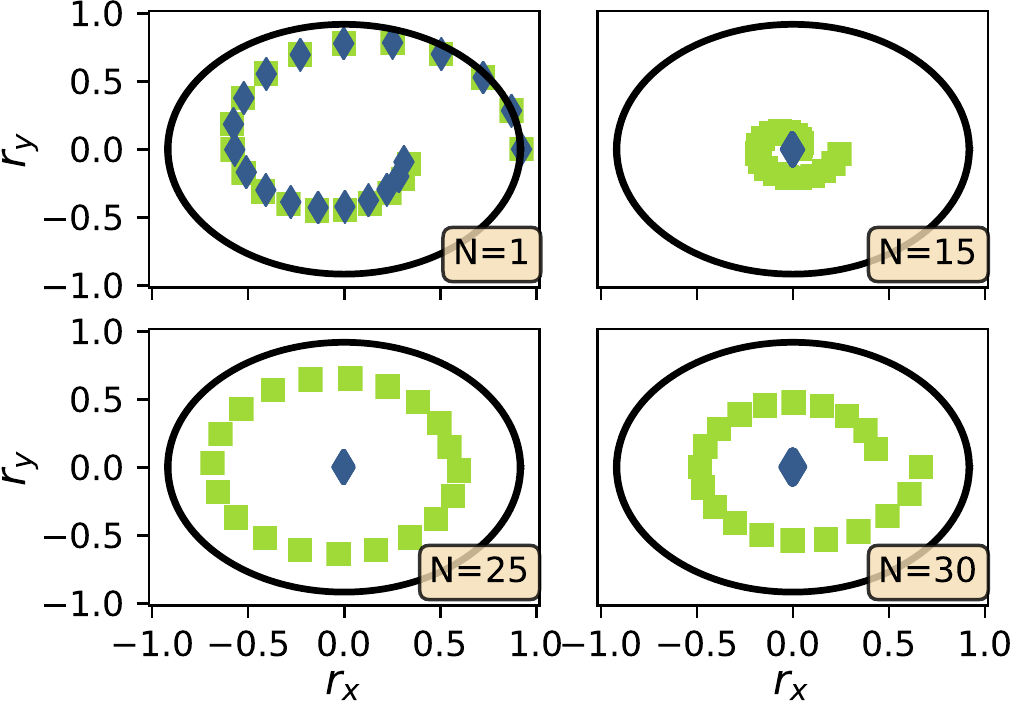}
\begin{footnotesize}
\caption{Each panel shows the trajectory of the probe along a specific period $N$ ($N=\tau \omega/2\pi$). The black solid line is for a unitary evolution where the trajectory of the probe is perfectly cyclic ($g=0$). The green squares refer to a non-unitary evolution caused by an integrable environmental chain ($h_z=0.0$) and the blue diamonds to a chaotic one ($h_z=0.5$). The initial state of the probe is a fixed pure state $(\theta=3\pi/8 ; \phi=0)$, R= 100 realizations for different pure random initial states for the environment were considered and the rest of the parameters are set as $\omega=1$, $h_x=1$, $J=1$, $g=0.2$, $L=9$. } 
\label{fig_geometric}
\end{footnotesize}
\end{center}
\end{figure}

Since we are dealing with a two-level-system acting as a probe, for a better geometrical understanding, in Fig. \ref{fig_geometric} we show the trajectories of the probe
parametrized by its Bloch vector $\vec{r}=(r_x,r_y,r_z)$ (recall that $r_z$ is constant) along specific periods for the same set of parameters of Fig. \ref{fig_dynamics}. As before, it can be seen that for short times ($N=1$) the trajectories in both regimes are indistinguishable. However, as the probe continues evolving for longer times, the decoherence produced by the chaotic environment is so strong that no revivals on the coherences are observed (in this regime, the trajectory of the probe is a quasi-stationary point). On the contrary, if the environment is integrable, there is a strong non-markovian behaviour that leads to huge revivals on both the coherences and the geometrical phase. For example, it can be observed from Fig. \ref{fig_geometric}  that despite along a specific period ($N=15$) the geometrical structure of the trajectory is almost entirely lost, several periods later ($N=25$) it has been quite well recovered with respect to the unitary evolution. This can be interpreted as a flow of information going from the probe to the environment during $N=15$ and as a backflow of information during $N=25$, which evidences a strong relation between non-Markovianity and the integrable nature of the environmental chain. (see Appendix B for more details)
\cite{breuer2009measure,rivas2014quantum,garcia2012non,mirkin2019information,mirkin2019entangling}. 

\renewcommand{\figurename}{Figure} 

Let us now extend our analysis. Until now, we have entirely focused on comparing the two limit cases of interest, i.e. a purely integrable regime with respect to a maximally chaotic one. But the important question is: can we extend our approach for reproducing the whole integrable to chaos transition when sweeping over a certain parameter range? To address this question, we should compare the standard spectral indicator $\eta$, which is already normalized, with the correction to the geometric phase $|\delta \Phi|$ (or some other dynamical quantity), which is not. Therefore, to normalize a given quantity $X$ when sweeping over a certain parameter $y$ on which $X$ depends, we can define 
\begin{equation}
X_{Norm}= \frac{\max (X(y))-X(y)}{\max (X(y))-\min(X(y))}.    
\end{equation}

\begin{figure}[!htb]
\begin{center}
\includegraphics[width=87mm]{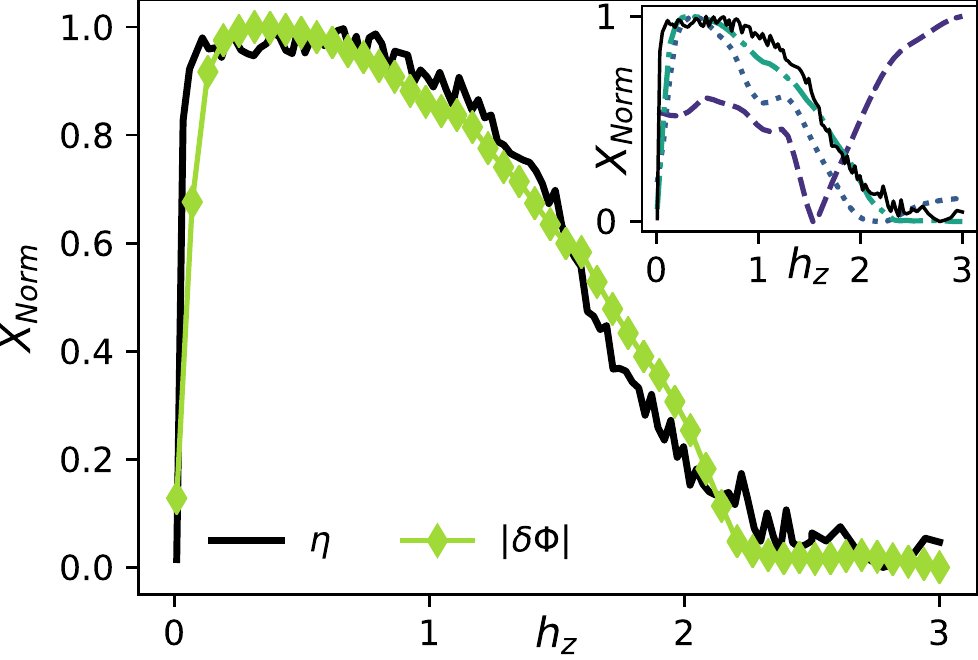}
\begin{footnotesize}
\caption{\textbf{Main panel:} Normalized correction $|\delta \Phi|$ averaged over 100 different realizations of pure random initial states for the environment as a function of $h_z$. The initial state of the probe is a fixed pure state $(\theta=3\pi/7; \phi=0)$, 
the environmental chain has a length of $L=9$ spins and $N=20$ periods are considered. The rest of the parameters are set as $\omega=1$, $h_x=1$, $J=1$ and $g=0.2$. Also, we plot the spectral indicator of chaos $\eta$ as a function of the parameter $h_z$. For computing $\eta$, a larger chain composed of $L=14$ spins was considered and only the odd subspace was taken into account ($D^{odd} \approx 8192$). \textbf{Inset panel:} Same as the main panel but for less periods: dashed line for $N=2$, dotted line for $N=5$ and dashdotted line for $N=15$, respectively. Same results were obtained for different initial states for the probe, i.e. using $\theta$ and $\phi$ as random angles for each realization. }  
\label{transition_ising}
\end{footnotesize}
\end{center}
\end{figure}

In Fig. \ref{transition_ising} we study the normalized correction $|\delta \Phi|$ averaged over 100 different realizations of pure random initial states for the environment, together with the spectral indicator of chaos $\eta$, both as a function of the parameter $h_z$. While for computing $\eta$ we have diagonalized a large spin chain of $L=14$ spins and taken into account the symmetries of the environment by analyzing only the odd subspace, for calculating  $|\delta \Phi|$ we have used a much shorter environmental spin chain ($L=9$) and its symmetries were completely neglected. Remarkably, we can see a strong correspondence between the degree of chaos present in the environmental chain (quantified with the standard spectral approach) with respect to the non-unitary correction on the accumulated geometric phase of the probe. However, this strong correspondence between the standard spectral characterization of quantum chaos and our decoherent approach holds only for sufficiently long times, as it is shown in the inset of Fig. \ref{transition_ising}. There can be seen that as we increase the number of periods involved in the dynamics, the degree of correction on the accumulated geometric phase of the probe converges to the degree of chaos present in the environmental spin chain being sensed.

\renewcommand{\figurename}{Figure} 

To gain further insight and test the universality of our decoherent mechanism for sensing quantum chaos, in the next subsections we will cover a wide range of physical systems with different conserved symmetries and interactions. We will show that this same approach holds in all those scenarios as well. 

%\onecolumngrid

\renewcommand{\figurename}{Figure} 
\begin{figure*}%[!htb]
\centering
\includegraphics[width=180mm]{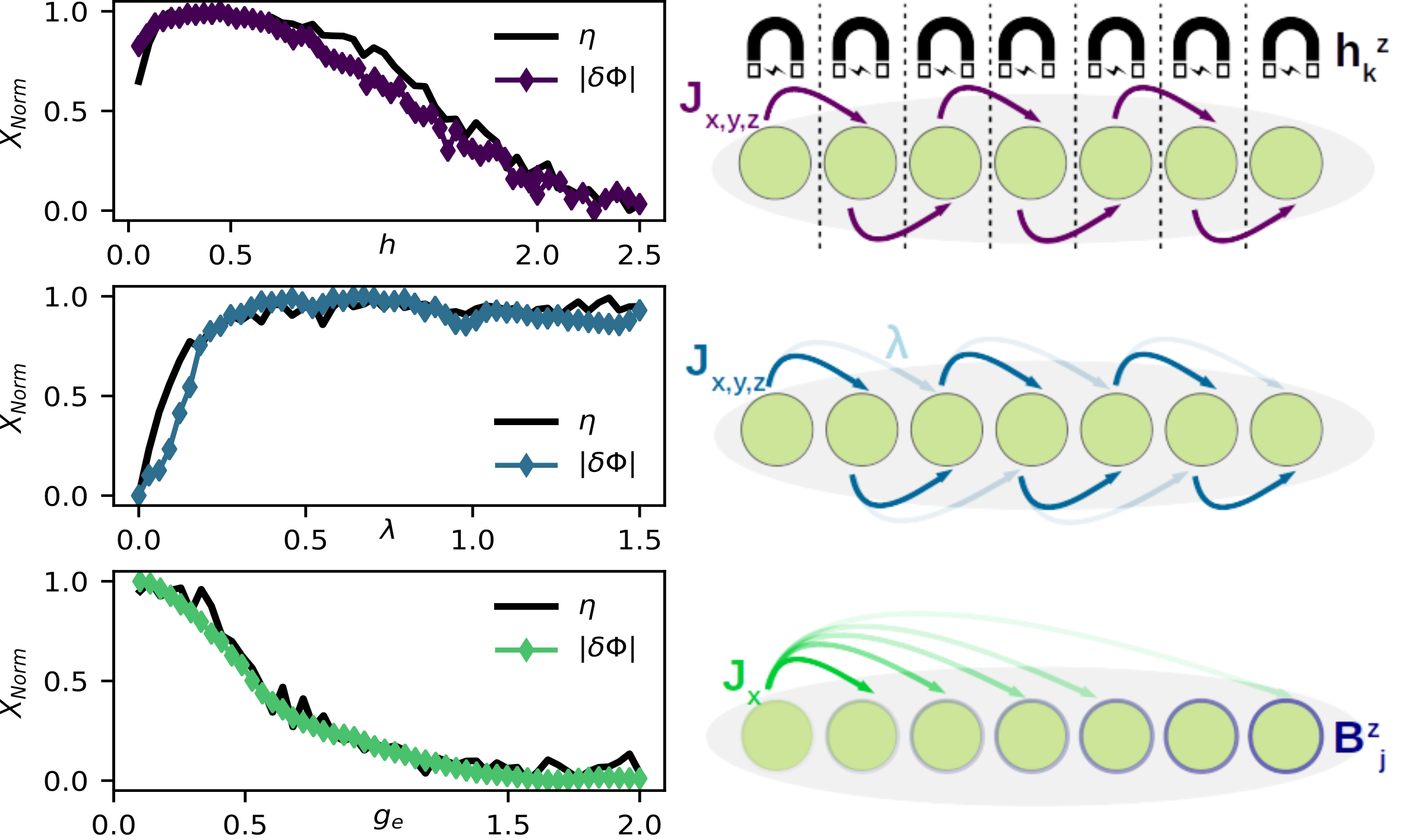}
\begin{footnotesize}
\caption{In all panels, we plot the normalized correction $|\delta \Phi|$ averaged over several different realizations of pure random initial states for the specific environmental chain under consideration, together with the spectral indicator of chaos $\eta$, as a function of a certain parameter which tunes the integrable to chaos transition in the environment. For computing $|\delta \Phi|$, the initial state of the probe is always the same fixed pure state $(\theta=3\pi/7; \phi=0)$,
the environmental chain has always a length of $L=9$ spins and $N=20$ periods are considered ($\omega=1$). \textbf{Panel (a): Heisenberg with random magnetic field.} The integrable to chaos transition is studied as a function of the random magnetic field strength $h$. For computing $\eta$, $50$ different realizations over random sets of $\{h_k^z\}$ were considered, a larger chain composed of $L=13$ spins was selected and only the subspace with $n=6$ was taken into account ($D_6 \approx 1716$). The coupling strength of the dephasing interaction was set as $g=0.005$. \textbf{Panel (b): Pertubed XXZ model.}  The integrable to chaos transition is studied as a function of
the perturbation with next-nearest-neighbor coupling $\lambda$. For computing $\eta$, a larger chain composed of $L=15$ spins was selected and only the subspace with $n=5$ was taken into account ($D_5 \approx 1500$). The rest of the parameters were set as $g=0.1$ and $\mu=0.5$.  \textbf{Panel (c): Ising with long range interactions.}  The integrable to chaos transition is studied as a function of the gradient strength $g_e$. The rest of the parameters were set as $J_0=1$, $B^{z0}=5$, $\gamma=1.3$ and $g=0.2$. For computing $\eta$, a larger chain composed of $L=13$ spins was considered and only the subspace with $n=6$ excitations was taken into account ($D_{6} \approx 1716$).} 
\label{transition_all}
\end{footnotesize}
\end{figure*}

%\twocolumngrid

\subsection{Heisenberg with random magnetic field}
The next environmental model to be sensed is a spin chain with nearest-neighbor interaction coupled to a random magnetic field in the $\hat{z}$ direction at each site. The Hamiltonian with open boundary conditions is given by 

\begin{equation}
\hat{H}_E=\sum_{k=1}^{L-1}\left(\hat{S}_{k}^{x}\hat{S}_{k+1}^{x} + \hat{S}_{k}^{y}\hat{S}_{k+1}^{y} + \hat{S}_{k}^{z}\hat{S}_{k+1}^{z} \right) + \sum_{k=1}^{L} h_k^z \hat{S}_{k}^{z},     
\label{heisenberg}
\end{equation}
where $\{h_k^z\}$ is a set of random variables uniformly distributed within the interval $[-h,h]$. In this model, the $\hat{z}$ component of the total spin  $\hat{S}^z=\sum_{k=1}^L \hat{S}_k^z$ is a conserved quantity. This conservation allows the separation of the spanned space into smaller subspaces $\hat{\mathcal{S}}_n$, where $n$ is a fixed quantity of spins up or down. The dimension of each subspace is given by 
\begin{equation}
    D_n= \dim\left(\hat{\mathcal{S}}_n\right)= \begin{pmatrix}
L\\
n 
\end{pmatrix}
=\frac{L!}{n!(L-n)!}.
\end{equation}
Taking into account this symmetry,  the system is integrable for $h=0$. But as the strength of the random magnetic field is increased, the degree of disorder is higher and the system reaches a Wigner-Dyson distribution near $h \simeq 0.5$. Finally, if the degree of disorder is too strong, there is a transition to a many-body localization (MBL) and the energy levels then follow a Poisson distribution again. In the upper panel of Fig. \ref{transition_all} we show the correction to the geometric phase of the probe $|\delta \Phi|$ averaged over 250 realizations of different  random sets of $\set{h_k^z}$ and different pure random initial states for the environment, together with the spectral chaos indicator $\eta$, both as a function of the random magnetic field $h$. As well as in the Ising spin model that has been analyzed before, for computing $|\delta \Phi|$ a shorter environmental chain was used and no considerations about its energy level symmetries were made. Despite this particular model presents disorder and a MBL, once again the global spectral characterization of quantum chaos coincides with our local decoherent approach for sufficiently long times.

\subsection{Perturbed XXZ model}
The many-body quantum system that we will sense now is an anisotropic spin chain with nearest-neighbor couplings and a perturbation with next-nearest-neighbor couplings. The Hamiltonian of this model with open boundary conditions is given by

\begin{equation}
\hat{H}_E(\lambda)=\hat{H}_0 + \lambda \hat{H}_1,
\end{equation}
where the parameter $\lambda$ quantifies the strength of the perturbation and each term is given by 
\begin{equation}
    \begin{split}
        & \hat{H}_0= \sum_{k=1}^{L-1}\left(\hat{\sigma}_{k}^{x}\hat{\sigma}_{k+1}^{x} + \hat{\sigma}_{k}^{y}\hat{\sigma}_{k+1}^{y} +\mu \hat{\sigma}_{k}^{z}\hat{\sigma}_{k+1}^{z} \right) \\
        & \hat{H}_1= \sum_{k=1}^{L-2}\left(\hat{\sigma}_{k}^{x}\hat{\sigma}_{k+2}^{x} + \hat{\sigma}_{k}^{y}\hat{\sigma}_{k+2}^{y} + \mu \hat{\sigma}_{k}^{z}\hat{\sigma}_{k+2}^{z} \right).
    \end{split}
\end{equation}

In this particular model, the integrable to chaos transition arises as the next-nearest-neighbor term becomes comparable to the nearest-neighbor one. Nevertheless, to observe the transition with the standard spectral measures, we must consider all the symmetries that this model presents. First, if the chain is isotropic ($\mu=1$), the total spin $\hat{S}^2$ is conserved, so we must consider a case with $\mu \neq 1$ to see the transition. Also, not only the $\hat{z}$ component of the spin $\hat{S}^z=\sum_{k=1}^L \hat{S}_k^z$ is a conserved quantity but also parity, so we need to analyze an odd or even subspace with a fixed number of excitations for being able to see the transition. In the middle panel of Fig. \ref{transition_all} we show the results for this model. The correction to the geometric phase of the probe $|\delta \Phi|$ as well as the standard spectral indicator $\eta$ are plotted as a function of the perturbation strength $\lambda$.
To sense homogeneously this environmental chain, for computing $|\delta \Phi|$ $100$ realizations were considered and a much shorter spin chain was used ($L=9$) with respect to the analysis regarding the spectral indicator of chaos $\eta$ ($L=15$).

\subsection{Ising with long range interactions}
The last model we will consider now differs from all others in the sense that it presents long-range interactions. The Hamiltonian is given by
\begin{equation}
    \hat{H}_E=\sum_{j<j^{\prime}} J_{j j^{\prime}} \sigma_{j}^{x} \sigma_{j^{\prime}}^{x}+\sum_{j=1}^{N}\left(B^{z 0}+(j-1) g_e \right) \sigma_{j}^{z},
\end{equation}
where the long-range interactions are given by
$J_{j j^{\prime}}=\frac{J_0}{|j-j^{\prime}|^\gamma}$, with $J_0$ being the nearest-neighbor coupling, $B^{z0}$ an overall bias field, $g_e$ a gradient strength and $\gamma$ the long range coefficient. We fix $\gamma=1.3$ and $B^{z0}/J_0 \gg 1 $ so that the total magnetization in the $\hat{z}$ direction is approximately conserved  \cite{morong2021observation}. In the lower panel of Fig. \ref{transition_all} we show the correction to the geometric phase of the probe $|\delta \Phi|$ averaged over 100 realizations of different pure random initial states for the environment, together with the spectral chaos indicator $\eta$, both as a function of $g_e$. Despite for computing $|\delta \Phi|$ a shorter chain was used and no considerations about the symmetries were made, as well as in all the other models, we are able to diagnosis quite accurately the degree of chaos through our local decoherent description.

\section{Conclusions}
In this work we have proposed a realistic decoherent mechanism for sensing the chaotic nature of a generic many-body quantum system. By locally coupling a two-level quantum probe to a many-body environmental chain through a dephasing interaction, we have shown that the long-time dynamics of the probe can be used as a sensor to monitor the degree of quantum chaos present in the chain. In particular, we have shown that the correction to the accumulated geometric phase of the probe with respect to its unitary evolution can be exploited as a robust tool for sensing the whole integrable to chaos transition within the specific spin chain that is acting as a many-body environment. For the sake of universality, we have used our decoherent mechanism for sensing the chaoticity on four different environmental spin chains, each one possessing different conserved symmetries and even in the presence of disorder and long-range interactions. Besides its universality, our method is experimentally friendly with current technologies since it does not require to consider long environmental chains neither to care about the energy level symmetries. By bridging the gap between Peres original idea of focusing on the long-time dynamics of the Loschmidt Echo, its relation to decoherence under pure dephasing interactions and the use of the non-unitary geometric phase as a sensor, we hope our findings to shed new light on the double nature of quantum chaos, both as a spectral and dynamical property.

\begin{acknowledgements}
The work of N.M. and D.W. was partially supported by CONICET (PIP 112201 50100493CO), UBACyT (20020130100406BA), and ANPCyT (PICT-2016-1056). P.I.V and F.C.L were supported by CONICET, UBACyT (20020170100252BA), and ANPCyT (PICT-2018-3801). P.I.V. acknowledges the International Centre of Theoretical Physics (ICTP) Associate Program. F.C.L. acknowledges Simons Associate Programme ICTP.
\end{acknowledgements}

%\clearpage

\appendix

\section{Averaged decoherence factor}
\label{appA}
\begin{figure}[!htb]
\begin{center}
\includegraphics[width=87mm]{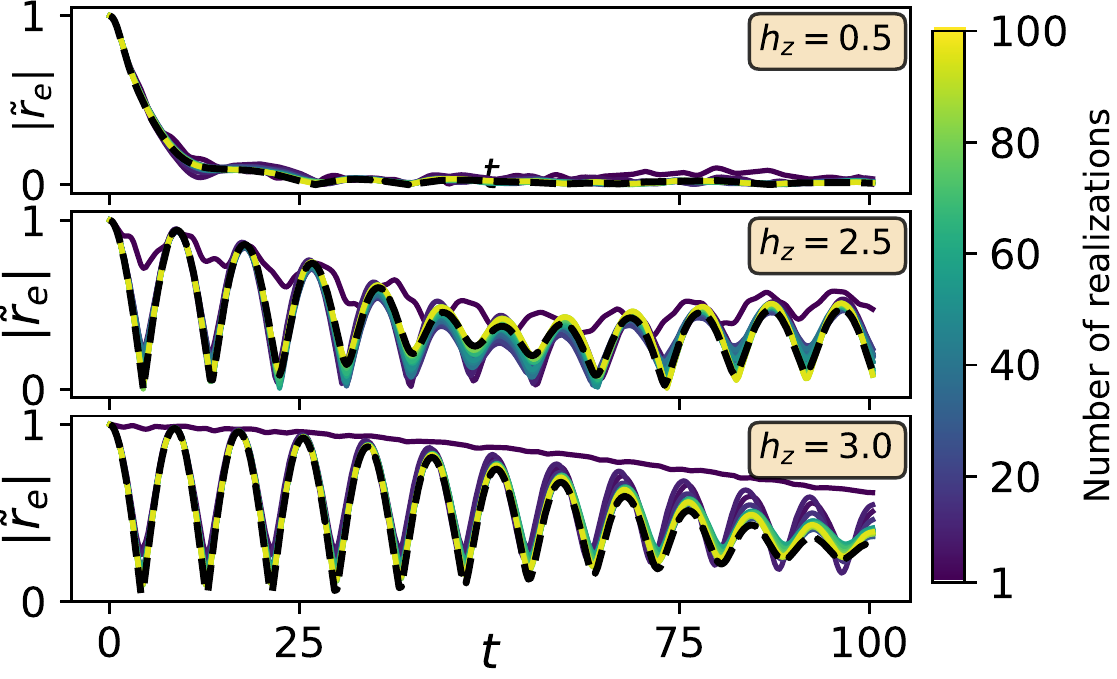}
\begin{footnotesize}
\caption{Absolute value of the averaged decoherence factor $|\tilde{r}_e(t)|$ as a function of time for different number of realizations over pure random initial states for the Ising environmental chain. As the number of realizations increases, the averaged decoherence factor $|\tilde{r}_e(t)|$ converges to the case of a maximally mixed initial state $\rho_\epsilon (0)=\mathbb{1}/2^L$, which is shown as a dashed black line. A different value of $h_z$ is considered for each panel and the rest of the parameters are set as $\omega=1$, $h_x=1$, $J=1$, $g=0.2$, $L=9$, $T=100$ (in units of $J^{-1}$).} 
\label{fig_convergence}
\end{footnotesize}
\end{center}
\end{figure}

The decoherence factor plays an important role whenever an open quantum system is coupled to an environment through a dephasing interaction. For convenience, it is usually assumed that the environment starts evolving from a specific pure state. Under such non-trivial assumption, the decoherence factor takes the form

\begin{equation}
        r(t)=  \bra{\varepsilon(0)} e^{i t\left[H_{E}- H_{\mathcal{S}\mathcal{E}} \right]} e^{-i t\left[H_{E}+ H_{\mathcal{S}\mathcal{E}} \right]}|\varepsilon(0)\rangle, 
\end{equation}
where $|\varepsilon(0)\rangle$ refers to the pure initial state of the environment. However, as we are interested in the task of sensing quantum chaos and this is a property of the entire spectrum of the environment, we cannot focus merely on one pure initial state. If we do this, great part of the environmental spectrum will not contribute to the dynamics of the probe
and thus we may be losing important information about the chaoticity of the environmental chain. This is the reason why we must introduce the concept of an averaged decoherence factor $\tilde{r}_e (t)$, where we simply take an average over different pure random initial states for the environment $|\varepsilon_m(0)\rangle$ (see Eq. (\ref{dec_factor_eff})). %, i.e. 
%\begin{equation}
%\begin{aligned}
%    \tilde{r}_{e}(t)&= \lim_{R\to\infty} \left[ \frac{1}{R}\sum_{m=1}^{R}  \bra{\varepsilon_m(0)} e^{i t\left[H_{E}- H_{\mathcal{S}\mathcal{E}} \right]} e^{-i t\left[H_{E}+ H_{\mathcal{S}\mathcal{E}} \right]}|\varepsilon_m(0) \rangle \right] .  
%\end{aligned}
%\end{equation}
As was argued in the main text, if the number of realizations $R$ is sufficiently large, given the full mixture of pure initial states involved in the average, $\tilde{r}_{e}(t)$ 
will converge to the decoherence factor associated to a maximally mixed initial state. This is precisely what is shown numerically in Fig. \ref{fig_convergence}
for the particular case of an environmental
Ising spin chain with transverse magnetic (see model in Eq. (\ref{ising})). However, the same argument holds for the rest of the models considered.

\renewcommand{\figurename}{Figure} 

Finally, it is also important to remark that the square of the absolute value of the decoherence factor is exactly what is known as the Loschmidt Echo (LE), i.e. $M(t)=|r(t)|^2$. In such a context, a common procedure is to average the LE over initial states according to the Haar measure, which is uniform over all quantum states in the Hilbert space. This averaged LE is given by
\begin{widetext}
\begin{equation}
\overline{M}(t)= \int d \ket{\varepsilon (0)} |r(t)|^2 = \frac{2^L+\left|\Tr(e^{i t\left[\hat{H}_{E}- \hat{H}_{\mathcal{S}\mathcal{E}} \right]} e^{-i t\left[\hat{H}_{E}+ \hat{H}_{\mathcal{S}\mathcal{E}} \right]} )\right|^2}{2^L(2^L+1)},
\end{equation}
\end{widetext}
which beyond a dimensional factor, presents the same qualitative behavior as the averaged decoherence factor $|\tilde{r}_e(t)|$ defined in Eq. (\ref{dec_factor_eff}).

\section{Chaos and non-Markovianity}
\label{appB}
Under the general framework of open quantum systems, non-Markovianity (NM) is usually associated with the existence of a flow of information that goes from the environment back to the open system. One standard way of quantifying the degree of NM under a given dynamics is to monitor the revivals of distinguishability \cite{medida1}. The distinguishability ($    \sigma(\hat{\rho}_{r1}(0),\hat{\rho}_{r2}(0),t)$) can be quantified by the derivative of the trace distance, which is defined as $\mathcal{D}(\hat{\rho}_{r1},\hat{\rho}_{r2})=\dfrac{1}{2}||\hat{\rho}_{r1}-\hat{\rho}_{r2}||$ and where $||A||=tr(\sqrt{A^{\dagger}A})$. In this context, under a Markovian regime the information of the open system is continuously leaked to the environment and thus quantum states become less and less distinguishable. On the contrary, under a non-Markovian dynamics, information can flow from the environment back to the reduced system and the distinguishability between states may increase within a given period of time. Consequently, a standard measure of NM (BLP measure \cite{medida1}) consists on integrating upon all the intervals of time where the distinguishability between two orthogonal initial states increases and maximizing over them i.e.

\begin{equation}
    \mathcal{N}^{BLP}=\max\limits_{{\lbrace \hat{\rho}_{r1}(0),\hat{\rho}_{r2}(0)\rbrace}} \int_{0, \sigma >0}^{\infty} \sigma \left (\hat{\rho}_{r1}(0),\hat{\rho}_{r2}(0),t'\right)  dt'.
    \label{BLP}
\end{equation}
Nevertheless, this measure of NM has some inconsistencies that have already been pointed out in the literature \cite{pineda2016measuring}. The main problem is that it overestimates the weight of fluctuations in the trace distance, which is an unfortunate fact due to the huge fluctuations that are present in our averaged decoherence factor in the integrable regime. To avoid this problem, a possible solution is to follow the approach proposed in Ref. \cite{pineda2016measuring}, where the calculation is restricted to the largest revival of $\mathcal{D}(\hat{\rho}_{r1},\hat{\rho}_{r2})$ with respect to its minimum value prior to the revival, instead of integrating upon all revivals, i.e 

\begin{equation}
\mathcal{N}^{LR}=\max_{t f, t \leq t f}\left[D\left(\hat{\rho}_{r1}\left(t_{f}\right), \hat{\rho}_{r2}\left(t_{f}\right)\right)-D\left(\hat{\rho}_{r1}(t), \hat{\rho}_{r2}(t)\right)\right].  
\end{equation}

\begin{figure}[!htb]
\begin{center}
\includegraphics[width=87mm]{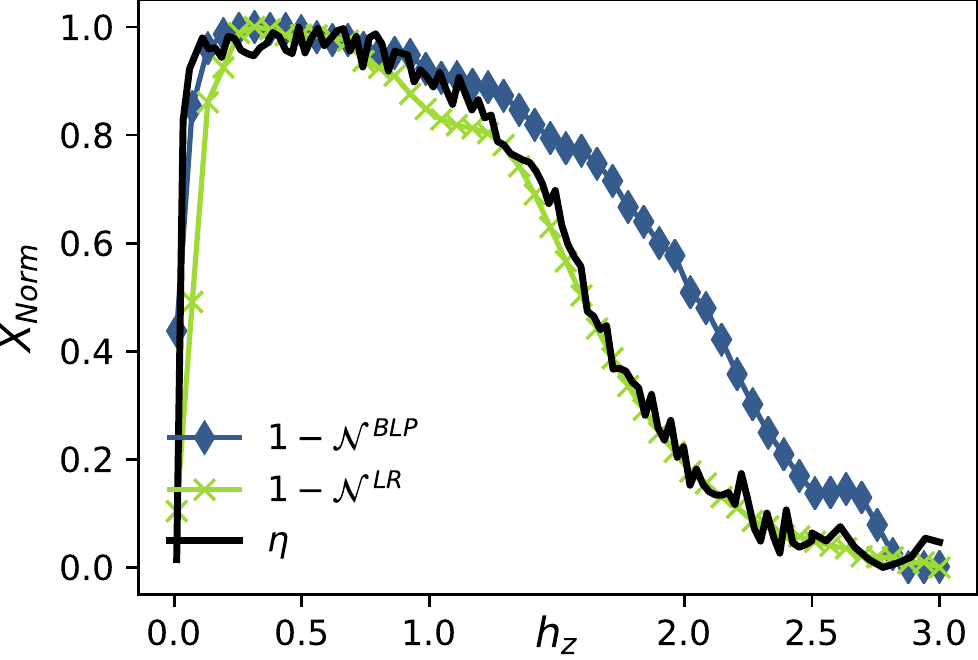}
\begin{footnotesize}
\caption{Normalized measures of NM ($\mathcal{N}^{BLP}_{Norm}$ and $\mathcal{N}^{LR}_{Norm}$) using the averaged decoherence factor $\tilde{r}_e (t)$ (see Eq. (\ref{dec_factor_eff})), as a function of $h_z$
. The pair of initial states that maximize both measures are $\hat{\rho}_{r1}(0)=\ket{+,x}\bra{+,x}$ ($\theta=\pi/2; \phi=0$) and $\hat{\rho}_{r2}(0)=\ket{-,x}\bra{-,x}$ ($\theta=\pi/2; \phi=\pi$). The environmental chain has a length of $L=9$ spins and $N=30$ periods are considered. The rest of the parameters are set as $\omega=1$, $h_x=1$, $J=1$ and $g=0.2$. Also, we plot the spectral indicator of chaos $\eta$ as a function of the parameter $h_z$. For computing $\eta$, a larger chain composed of $L=14$ spins was considered and only the odd subspace was taken into account ($D^{odd} \approx 8192$).} 
\label{no_marko}
\end{footnotesize}
\end{center}
\end{figure}

In Fig. \ref{no_marko} we study the relation between quantum chaos (quantified through the standard spectral indicator $\eta$) and both measures of NM  ($\mathcal{N}^{BLP}_{Norm}$ and $\mathcal{N}^{LR}_{Norm}$) as a function of $h_z$, for the particular case of an environmental Ising spin chain with transverse magnetic (see model in Eq. (\ref{ising})). Remarkably, we can see a great degree of correspondence between the chaotic nature of the environment and the $\mathcal{N}^{LR}$ measure of NM. In particular, it is clear that the more chaotic the environment is, the dynamics is more Markovian ($\mathcal{N}^{LR}_{Norm} \simeq 0$). On the contrary, if the environment is integrable, the degree of NM is higher. This is consistent with the fact that the correction to the geometric phase of the probe in the integrable regime is lower with respect to the chaotic regime. While there is a significant backflow of information in the integrable case allowing the probe to partially recover the information that was previously lost, this is not the case in the chaotic regime and thus the correction is much higher. This constitutes another concrete example where NM can be exploited as a resource for quantum information protocols \cite{mirkin2019entangling,mirkin2019information,mirkin2020quantum,bylicka2013non,berk2019resource,anand2019quantifying,bhattacharya2020convex,env_resource0}. Moreover, since NM is explicitly defined for the long-time regime ($T \to \infty$) and we are obtaining a remarkable correspondence between the degree of NM in the dynamics and the degree of chaos in the environment, this is another argument for the fact that quantum chaos is a spectral property that manifests itself in the long-time dynamics.

\renewcommand{\figurename}{Figure} 

%\clearpage
\bibliography{main.bib}
\end{document}